\documentclass[12pt]{article}
\usepackage{amssymb,amsmath,epsfig}

\begin{document}

\title{\bf Warm Inflation in $f(\mathcal{G})$ Theory of Gravity}
\author{M. Sharif \thanks {msharif.math@pu.edu.pk} and Ayesha Ikram
\thanks{ayeshamaths91@gmail.com}\\
Department of Mathematics, University of the Punjab,\\
Quaid-e-Azam Campus, Lahore-54590, Pakistan.}

\date{}
\maketitle

\begin{abstract}
The aim of this paper is to explore warm inflation in the background
of $f(\mathcal{G})$ theory of gravity using scalar fields for the
FRW universe model. We construct the field equations under slow-roll
approximations and evaluate the slow-roll parameters, scalar and
tensor power spectra and their corresponding spectral indices using
viable power-law model. These parameters are evaluated for a
constant as well as variable dissipation factor during intermediate
and logamediate inflationary epochs. We also find the number of
e-folds and tensor-scalar ratio for each case. The graphical
behavior of these parameters proves that the isotropic model in
$f(\mathcal{G})$ gravity is compatible with observational Planck
data.
\end{abstract}

\section{Introduction}

Recent cosmological observations discover revolutionary features of
the present universe and deduce that the universe is experiencing a
uniform accelerated expansion. Experimental data from supernova type
Ia, cosmic microwave background radiation (CMBR), large scale
structure (LSS) etc provide evidences for this cosmic acceleration
\cite{1}. This increasing rate of cosmic expansion is a consequence
of mysterious force called dark energy (DE) supposed to have large
negative pressure. Modified theories of gravity are the most
promising approaches to explore the nature of DE. These
modifications are obtained by adding or replacing the curvature
invariants or their corresponding generic functions. These gravity
theories include $f(R)$ theory ($R$ is the Ricci scalar),
Brans-Dicke theory, Gauss-Bonnet (GB) theory etc \cite{2}.

Gauss-Bonnet invariant $(\mathcal{G})$ is defined as a linear
combination of $R^{2}$, the Ricci tensor $(R_{\mu\nu})$ and the
Riemann tensor $(R_{\mu\nu\tau\xi})$. It has non-trivial
contribution to the field equations for dimensions $\geq5$ while it
is four dimensional topological term. In order to study the
contribution of $\mathcal{G}$ in four dimensions, Nojiri and
Odintsov \cite{3} introduced a new modified theory of gravity by
adding an arbitrary function $f(\mathcal{G})$ in the
Einstein-Hilbert action. This theory is named as modified GB theory
or $f(\mathcal{G})$ theory of gravity. Modified GB theory is another
alternative to discuss DE and efficiently describes the late-time
cosmic acceleration for the effective equation of state, transition
from deceleration to acceleration as well as passes the solar system
tests \cite{3,4}. De Felice and Tsujikawa \cite{4a} studied the
solar system constraints on cosmologically viable $f(\mathcal{G})$
gravity models that are responsible for late-time cosmic
acceleration and found that these models are consistent with solar
system constraints for a wide range of model parameters.

Apart from current cosmic expansion, the universe also went through
a rapid expansion in the early time named as inflationary era.
Inflation is the natural solution to cure shortcomings of standard
model of cosmology (big-bang model) such as the horizon, flatness,
monopole etc \cite{6}. The origin of anisotropies observed in CMBR
is explained elegantly by this early era and provides a fascinating
mechanism to interpret LSS of the universe. Scalar field acts as
source of inflation (inflaton) which is a combination of potential
and kinetic terms coupled with gravity. There are two phases of
inflationary regime. Firstly, the universe inflates and scalar field
interactions with other fields become worthless (slow-roll). In this
evolutionary stage, the potential energy dominates over kinetic
energy \cite{8}. The other is the reheating phase in which inflaton
decays into matter and radiations. This is the end stage of
inflationary epoch where both energies are comparable and the
inflaton starts to swing about minimum potential \cite{9}.

The most appealing challenge for researchers is how to connect the
universe towards the end of inflationary era. Berera \cite{10} gave
the idea of warm inflation opposite to cold which unifies the
slow-roll and reheating phases. During the slow-roll regime, the
inflaton field is decomposed into matter and radiations. Dissipation
effects are of significant importance in this inflationary epoch
that arise from the friction term. The thermal fluctuations play a
dominant role in the production of initial density fluctuations
required for LSS formation. The vacuum energy converts into
radiation energy during inflationary period and thus smoothly enters
into the radiation dominated regime \cite{11}.

There are several forms of scale factors like intermediate and
logamediate scenarios which are used to discuss the inflationary
era. Intermediate inflation is originated from the string theory and
is faster than the power-law inflation but slower than the de Sitter
\cite{12}. The concept of logamediate appeared in scalar-tensor
theories \cite{13}. Herrera et al. \cite{14} studied general
dissipative coefficient in these regimes and analyzed them in both
strong and weak dissipative regimes. Setare and Kamali \cite{15}
explored warm inflation using vector fields for FRW universe model
in intermdiate as well as logamediate inflationary epochs and found
consistent results with WMAP7. Sharif and Saleem \cite{16} proved
that locally rotationally symmetric Bianchi I universe model is
compatible with WMAP7 in the context of warm vector inflation.

Inflation has also become a debatable issue in modified theories of
gravity. Banijamali and Fazlpour \cite{16a} discussed the power-law
inflation in non-minimal Yang-Mills $f(\mathcal{G})$ gravity in the
background of Einstein as well as $f(\mathcal{G})$ gravity and
proved that such theories explain both inflation as well as
late-cosmic acceleration. Bamba et al. \cite{17} investigated the
parameters of inflationary models in the framework of $f(R)$ gravity
through the reconstruction methods. They concluded that several
$f(R)$ models, especially, a power-law model gives the best fit
values in agreement with BICEP2 and Planck results. Bamba and
Odintsov \cite{18} studied inflationary cosmology in $R^2$ gravity
with its extensions to generalize the Starobinsky inflationary
model. It is found that the spectral index of scalar modes of
density perturbations and the tensor-scalar ratio are consistent
with the Planck results. De Laurentis et al. \cite{18a} studied
cosmological inflation in $f(R,\mathcal{G})$ gravity by considering
two effective masses in which one is related to $R$ and other is
related to $\mathcal{G}$. These corresponding masses discussed the
dynamics at early and very early epochs of the universe, giving rise
to a natural double inflationary scenario.

In this paper, we explore warm inflation driven by scalar fields in
$f(\mathcal{G})$ gravity for isotropic and homogeneous universe
model. The paper has the following format. In section \textbf{2}, we
construct the field equations and discuss warm inflationary
dynamics. Section \textbf{3} deals with constant and variable
dissipation factors for intermediate regime. We evaluate the
slow-roll parameters, scalar, tensor power spectra and their
corresponding spectral indices. In section \textbf{4}, the same
parameters are calculated for logamediate inflation. We conclude the
results in the last section.

\section{Warm Inflationary Dynamics}

In this section, we formulate the field equations and discuss
dynamics for warm inflation. The universe is filled with radiation
and self-interacting scalar fields. The Lagrangian is given by
\cite{19}
\begin{equation}\label{1}
\mathcal{I}=\int
d^{4}x\sqrt{-g}\left(\frac{R}{2\kappa^2}+f(\mathcal{G})
-\frac{1}{2}g^{\mu\nu}\partial_{\mu}\varphi\partial
_{\nu}\varphi-U(\varphi)\right),
\end{equation}
where $\kappa^2=\frac{8\pi}{M_{Pl}^2}$ is the coupling constant in
which $M_{Pl}^2$ is the Planck mass and
$\mathcal{G}=R^2-4R_{\mu\nu}R^{\mu\nu}+R_{\mu\nu\tau\xi}R^{\mu\nu\tau\xi}$.
The curvature terms represent gravitational part of the Lagrangian
whereas matter part corresponds to scalar field and the potential
function ($U(\varphi)$). The line element for FRW universe model is
\begin{equation}\label{2}
ds^2=-dt^2+a^2(t)(dx^2+dy^2+dz^2),
\end{equation}
where $a(t)$ denotes the scale factor depending on cosmic time. The
Ricci scalar and GB invariant for Eq.(\ref{2}) are
\begin{equation}\nonumber
R=12H^2+6\dot{H}, \quad \mathcal{G}=24H^2(H^2+\dot{H}),
\end{equation}
where $H=\frac{\dot{a}}{a}$ is the Hubble parameter and dot being
the derivative with respect to $t$. In warm inflation, we consider
that the total energy density of the universe is the sum of energy
density associated with scalar $(\rho_{\varphi})$ and radiation
field $(\rho_{r})$. The corresponding field equations for perfect
fluid are
\begin{eqnarray}\label{9}
\rho_{\varphi}+\rho_{r}&=&\frac{3H^2}{\kappa^2}
+f(\mathcal{G})-\mathcal{G}f_{\mathcal{G}}
+24H^3\dot{\mathcal{G}}f_{\mathcal{GG}},\\\nonumber
P_{\varphi}+P_{r}&=&-\left(\frac{1}{\kappa^2}(3H^2+2\dot{H})
+f(\mathcal{G})-\mathcal{G}f_{\mathcal{G}}
+16(H\dot{H}+H^3)\dot{\mathcal{G}}f_{\mathcal{GG}}\right.
\\\label{9a}&+&\left.8H^2\ddot{\mathcal{G}}f_{\mathcal{GG}}
+8H^2\dot{\mathcal{G}}^2f_{\mathcal{GGG}}\right),
\end{eqnarray}
where $f_{\mathcal{G}}=\frac{df}{d\mathcal{G}}$, $P_{\varphi}$ and
$P_{r}$ are the pressures of scalar and radiation fields,
respectively. The energy density and pressure for scalar field are
found as
\begin{equation}\nonumber
\rho_{\varphi}=\frac{1}{2}\dot{\varphi}^2+U(\varphi), \quad
P_{\varphi}=\frac{1}{2}\dot{\varphi}^2-U(\varphi).
\end{equation}
The dynamical equations are described by
\begin{eqnarray}\label{10}
\dot{\rho}_{\varphi}+3H(\rho_{\varphi}+P_{\varphi})&=&
-\chi\dot{\varphi}^2,\\\label{11}
\dot{\rho}_{r}+4H\rho_{r}&=&\chi\dot{\varphi}^2,
\end{eqnarray}
where $\chi>0$ is the friction or dissipation factor. It describes
the decay of inflaton into radiation during inflationary epoch.
Dissipation factor may be a function of scalar field
$\chi(\varphi)$, the temperature of thermal bath $\chi(T)$, both
$\chi(\varphi,T)$ or a constant. A general form of dissipation
coefficient is \cite{19a}
\begin{equation}\nonumber
\chi(\varphi,T)=\chi_{*}\frac{T^m}{\varphi^{m-1}},
\end{equation}
where $\chi_{*}$ is a constant associated with dissipative
microscopic dynamics and $m$ is an integer. Different expressions
for $\chi$ are obtained for different values of $m$. When $m=-1$,
$\chi\propto\frac{\varphi^{2}}{T}$ which corresponds to the
non-supersymmetry (SUSY) case whereas $m=0$ yields
$\chi\propto\varphi$ that associated with exponentially decaying
propagator in the SUSY case. For $m=1$, $\chi\propto T$ represents
the high-temperature SUSY case and the value of $m=3$ implies
$\chi\propto\frac{T^3}{\varphi^2}$ which corresponds to
low-temperature case \cite{19b}. In this work, we consider the
following two forms of $\chi$ \cite{15}:
\begin{itemize}
\item $\chi=\chi_{0}=$ constant;
\item $\chi=\chi_{*}\frac{T^3}{\varphi^2}$.
\end{itemize}

During warm inflation, $\rho_{\varphi}$ dominates over $\rho_{r}$
and the radiation production is quasi-stable where
\begin{equation}\label{12}
\dot{\rho}_{r}\ll 4H\rho_{r}, \quad
\dot{\rho}_{r}\ll\chi\dot{\varphi}^2.
\end{equation}
Since kinetic energy is negligible with respect to potential, so we
have $P_{\varphi}=-\rho_{\varphi}$. The slow-roll approximations are
\begin{equation}\label{13}
\dot{\varphi}^2\ll U(\varphi),\quad
\ddot{\varphi}\ll(3H+\chi)\dot{\varphi}.
\end{equation}
Using the conditions (\ref{12}) and (\ref{13}), Eqs.(\ref{10}) and
(\ref{11}) reduce to
\begin{eqnarray}\label{14}
3H(1+\mathcal{R})\dot{\varphi}&=&-U'(\varphi),
\\\label{15}\rho_{r}=\frac{\chi\dot{\varphi}^2}{4H}
&=&\frac{3}{4}\mathcal{R}\dot{\varphi}^2=CT^4,
\end{eqnarray}
where prime denotes derivative with respect to
$\varphi,~C=\frac{\pi^2\tilde{g}}{30}$ ($\tilde{g}$ represents the
number of relativistic degrees of freedom) and
$\mathcal{R}=\frac{\chi}{3H}$ is called the decay or dissipation
rate. The value of $\mathcal{R}$ is greater than $1$ in strong
dissipative region and less than $1$ in weak region. The slow-roll
parameters $(\varepsilon,\eta)$ are defined as
\begin{equation}\label{16}
\varepsilon=-\frac{\dot{H}}{H^2},\quad
\eta=-\frac{\ddot{H}}{H\dot{H}},
\end{equation}
where $\left|\frac{\dot{H}}{H^2}\right|\ll1$ and
$\left|\frac{\ddot{H}}{H\dot{H}}\right|\ll1$ during inflation. The
time derivative of Eq.(\ref{9}) with these conditions yield
\begin{equation}\label{17}
\frac{6H\dot{H}}{\kappa^2}-2304\dot{H}H^7f_{\mathcal{GG}}\simeq
U'(\varphi)\dot{\varphi}.
\end{equation}
Equations (\ref{14}) and (\ref{17}) give
\begin{equation}\label{18}
\dot{\varphi}^2=-\frac{2\dot{H}}{(1+\mathcal{R})}\left[\frac{1}{\kappa^2}
-384H^6f_{\mathcal{GG}}\right].
\end{equation}
The temperature of thermal bath is obtained by using Eq.(\ref{18})
in (\ref{15}) as
\begin{equation}\label{19}
T=\left[-\frac{3\mathcal{R}\dot{H}}{2C(1+\mathcal{R})}\left(\frac{1}
{\kappa^2}-384H^6f_{\mathcal{GG}}\right)\right]^{\frac{1}{4}}.
\end{equation}

Now, we calculate perturbations for FRW universe model by the
variation of field $\varphi$. In non-warm and warm inflationary
scenarios, the fluctuations of $\varphi$ are obtained by quantum and
thermal fluctuations, respectively as
\begin{equation}\label{20}
<\delta\varphi>_{quantum}=\frac{H}{2\pi},\quad
<\delta\varphi>_{thermal}=\left(\frac{\chi
HT^2}{(4\pi)^3}\right)^{\frac{1}{4}}.
\end{equation}
The scalar power spectrum $(\mathcal{P}_{s})$ and its spectral index
$(n_{s})$ are defined as \cite{20}
\begin{eqnarray}\label{21}
\mathcal{P}_{s}&=&\left(\frac{H}{\dot{\varphi}}<\delta\varphi>\right)^2,
\\\label{22}n_{s}&=&1+\frac{d\ln\mathcal{P}_{s}}{d\ln k}.
\end{eqnarray}
Using Eqs.(\ref{18}) and (\ref{20}) in (\ref{21}), the expression
for $\mathcal{P}_{s}$ with strong dissipative region leads to
\begin{equation}\label{23}
\mathcal{P}_{s}=-\frac{H^{\frac{3}{2}}}{6\dot{H}}
\left[\frac{1}{\kappa^2}-384H^6f_{\mathcal{GG}}\right]^{-1}
\left(\frac{\chi^3T^2}{(4\pi)^3}\right)^{\frac{1}{2}}.
\end{equation}
For the tensor perturbations, we have
\begin{eqnarray}\label{24}
\mathcal{P}_{T}&=&8\kappa^2\left(\frac{H}{2\pi}\right)^2,
\\\label{25}n_{T}&=&-2\varepsilon,
\end{eqnarray}
where $\mathcal{P}_{T}$ and $n_{T}$ represent the tensor power
spectrum and tensor spectral index, respectively. The tensor-scalar
ratio in $f(\mathcal{G})$ theory becomes
\begin{equation}\label{26}
r=-\left(\frac{144\kappa^4(4\pi)^3}{\chi^3\pi^4T^2}\right)^{\frac{1}{2}}
\left[\frac{1}{\kappa^2}-384H^6f_{\mathcal{GG}}\right]H^{\frac{1}{2}}\dot{H}.
\end{equation}
According to recent observations from Planck data \cite{22}, the
scalar spectral index is constrained to
$n_{s}=0.9603\pm0.0073(68\%CL)$ while $r<0.11(95\%CL)$ is the
physical acceptable range showing the expanding universe.

\section{Intermediate Inflation}

In this section, we discuss warm intermediate inflation for the
power-law model as
\begin{equation}\label{27}
f(\mathcal{G})=\alpha\mathcal{G}^n, \quad n>1,
\end{equation}
where $\alpha$ is an arbitrary constant. During inflation, the value
of $f(\mathcal{G})$ for $n>1$ dominates over the Einstein-Hilbert
term \cite{19}. Intermediate inflation is motivated by string theory
and proved to be the exact solution of inflationary cosmology
containing a particular form of the scale factor. In this era, the
universe expands at the rate slower than the standard de Sitter
inflation (with scale factor $a(t)=a_{0}\exp{(H_{0}t)}$) while
faster than power-law inflation (with scale factor
$a(t)=t^{p},~p>1$). During this regime, the scale factor evolves as
\cite{12}
\begin{equation}\label{28}
a(t)=a_{0}\exp(\gamma t^{g}),\quad \gamma>0, \quad0<g<1.
\end{equation}
The corresponding number of e-folds is given by
\begin{equation}\label{29}
\mathcal{N}=\int_{t_{i}}^{t}Hdt= \gamma(t^{g}-t_{i}^{g}),
\end{equation}
where $t_{i}$ is the beginning time of inflationary epoch. In the
following, we discuss dynamics of warm inflation for constant as
well as variable dissipation factor.

\subsection{Case I: $\chi=\chi_{0}$}

Here, we calculate the parameters discussed in the previous section.
Using Eqs.(\ref{18}), (\ref{27}) and (\ref{28}), the solutions of
inflaton and Hubble parameter give
\begin{equation}\label{30}
\varphi=\varphi_{0}+\Delta_{1} t^{\frac{1}{2}[4n(g-1)+1]}, \quad
H=\gamma g\left(\frac{\varphi-\varphi_{0}}
{\Delta_{1}}\right)^{\frac{2(g-1)}{4n(g-1)+1}},
\end{equation}
where $\Delta_{1}=\left[\frac{384(24)^{n-1}\alpha n(n-1)(\gamma
g)^{4n}(g-1)}{\chi_{0}[4n(g-1)+1]^2}\right]^{\frac{1}{2}},~\alpha$
must be negative to make $\Delta_{1}$ real. The slow-roll parameters
become
\begin{eqnarray}\label{31}
\varepsilon&=&\left(\frac{1-g}{\gamma
g}\right)\left(\frac{\varphi-\varphi_{0}}{\Delta_{1}}\right)^
{\frac{-2g}{4n(g-1)+1}},\\\label{32}\eta&=&\left(\frac{2-g}{\gamma
g}\right)\left(\frac{\varphi-\varphi_{0}}{\Delta_{1}}\right)
^{\frac{-2g}{4n(g-1)+1}}.
\end{eqnarray}
Using Eqs.(\ref{18}), (\ref{27}), (\ref{28}) and (\ref{30}) in
(\ref{15}), the radiation density takes the form
\begin{equation}\nonumber
\rho_{r}=\frac{3}{2}384(24)^{n-2}\alpha n(n-1)(\gamma
g)^{4n-1}(g-1)\left(\frac{\varphi-\varphi_{0}}{\Delta_{1}}\right)
^{\frac{2[4n(g-1)-g]}{4n(g-1)+1}}.
\end{equation}
The number of e-folds in terms of scalar field is calculated from
Eqs.(\ref{29}) and (\ref{30}) as
\begin{equation}\label{33}
\mathcal{N}=\gamma\left[\left(\frac{\varphi-\varphi_{0}}{\Delta_{1}}
\right)^{\frac{2g}{4n(g-1)+1}}-\left(\frac{\varphi_{i}-\varphi_{0}}
{\Delta_{1}}\right)^{\frac{2g}{4n(g-1)+1}}\right].
\end{equation}
The initial inflaton magnitude $\varphi_{i}$ is found by fixing
$\varepsilon=1$ as
\begin{equation}\label{34}
\varphi_{i}=\varphi_{0}+\Delta_{1}\left(\frac{1-g}{\gamma
g}\right)^{\frac{4n(g-1)+1}{2g}}.
\end{equation}
Using this value of $\varphi_{i}$ in Eq.(\ref{33}), we get $\varphi$
in terms of $\mathcal{N}$ as
\begin{equation}\label{35}
\varphi=\varphi_{0}+\Delta_{1}\left(\frac{\mathcal{N}}{\gamma}+\frac{1-g}{\gamma
g}\right)^{\frac{4n(g-1)+1}{2g}}.
\end{equation}

The scalar power spectrum can be calculated using Eqs.(\ref{19}),
(\ref{27}), (\ref{28}) and (\ref{30}) in (\ref{23}) as follows
\begin{eqnarray}\nonumber
\mathcal{P}_{s}&=&\left(\frac{\chi_{0}^{3}}{36(4\pi)^3}\right)^{\frac{1}{2}}
\left(\frac{3}{2C}\right)^{\frac{1}{4}}[384(24)^{n-2}\alpha
n(n-1)(\gamma g)^{4n-3}(g-1)]^{\frac{-3}{4}}\\\label{36}
&\times&\left(\frac{\varphi-\varphi_{0}}{\Delta_{1}}\right)
^{\frac{12n-12ng+9g-6}{8n(g-1)+2}}.
\end{eqnarray}
With the help of Eq.(\ref{35}), it can also be written in terms of
$\mathcal{N}$ as
\begin{eqnarray}\nonumber
\mathcal{P}_{s}&=&\left(\frac{\chi_{0}^{3}}{36(4\pi)^3}\right)^{\frac{1}{2}}
\left(\frac{3}{2C}\right)^{\frac{1}{4}}[384(24)^{n-2}\alpha
n(n-1)(\gamma g)^{4n-3}(g-1)]^{\frac{-3}{4}}\\\label{37}
&\times&\left(\frac{\mathcal{N}}{\gamma}+\frac{1-g}{\gamma
g}\right)^{\frac{12n-12ng+9g-6}{4g}}.
\end{eqnarray}
Using Eqs.(\ref{36}) and (\ref{37}) in (\ref{22}), the following
expressions for $n_{s}$ are obtained
\begin{eqnarray}\nonumber
n_{s}&=&1-\left(\frac{12n-12ng+9g-6}{4\gamma g}\right)
\left(\frac{\varphi-\varphi_{0}}{\Delta_{1}}\right)^{\frac{-2f}{4n(g-1)+1}},
\\\label{38}&=&1-\left(\frac{12n-12ng+9g-6}{4\gamma g}\right)
\left(\frac{\mathcal{N}}{\gamma}+\frac{1-g}{\gamma g}\right)^{-1}.
\end{eqnarray}
Figure \textbf{1} shows the graphical behavior of $n_{s}$ against
number of e-folds. In the left plot, the observational value of
$n_{s}$ corresponds to $\mathcal{N}=64,~94$ and $133$ for
$n=1.1,~1.5$ and $2$, respectively which shows that $\mathcal{N}$
has a direct relation with $n$. Similarly, for the right plot, the
number of e-folds varies from $36,~48,~64$ and $104$ for
$n=1.5,~2,~3$ and $5$, respectively. Figure \textbf{1} indicates
that e-folds reduces as $g$ approaches to $1$.
\begin{figure}
\epsfig{file=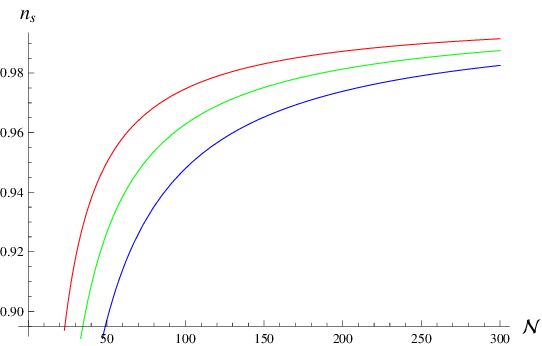, width=0.5\linewidth}\epsfig{file=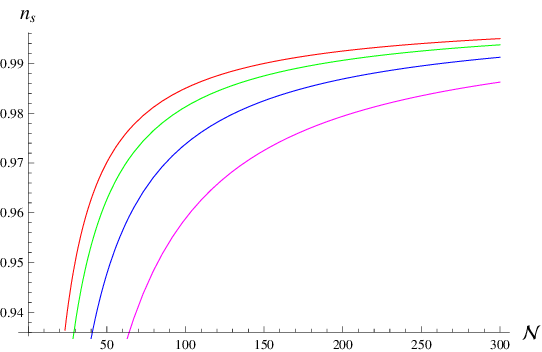,
width=0.5\linewidth}\caption{$n_{s}$ versus $\mathcal{N}$ for
$\gamma=1$. The left plot is for $g=0.5,~n=1.1$ (red), $n=1.5$
(green) and $n=2$ (blue) and right for $g=0.8,~n=1.5$ (red), $n=2$
(green), $n=3$ (blue) and $n=5$ (magenta).}
\end{figure}

The tensor power spectrum is obtained using Eqs.(\ref{24}),
(\ref{30}) and (\ref{35}) as
\begin{eqnarray}\nonumber
\mathcal{P}_{T}&=&\frac{2\kappa^2}{\pi^2}(\gamma
g)^2\left(\frac{\mathcal{N}}{\gamma}+\frac{1-g}{\gamma
g}\right)^{\frac{2}{g}(g-1)}.
\end{eqnarray}
Equations (\ref{25}) and (\ref{31}) give the expression for tensor
spectrum as
\begin{eqnarray}\nonumber
n_{T}=2\frac{(g-1)}{\gamma
g}\left(\frac{\mathcal{N}}{\gamma}+\frac{1-g}{\gamma g}\right)^{-1}.
\end{eqnarray}
Using Eqs.(\ref{19}), (\ref{26}), (\ref{29}), (\ref{30}) and
(\ref{35}), the tensor-scalar ratio is given by
\begin{eqnarray}\nonumber
r&=&\left(\frac{144\kappa^4(4\pi)^3}{\chi_{0}^3\pi^4}\right)^{\frac{1}{2}}
\left(\frac{2C}{3}\right)^{\frac{1}{4}}(384(24)^{n-2}\alpha
n(n-1)(g-1))^{\frac{3}{4}}(\gamma
g)^{\frac{1}{4}(12n-1)}\\\nonumber&\times&\left(\frac{\mathcal{N}}{\gamma}
+\frac{1-g}{\gamma g}\right)^{\frac{(12ng-12n-g-2)}{4g}}.
\end{eqnarray}
In terms of $n_{s}$, the above equation becomes
\begin{eqnarray}\nonumber
r&=&\left(\frac{144\kappa^4(4\pi)^3}{\chi_{0}^3\pi^4}\right)^{\frac{1}{2}}
\left(\frac{2C}{3}\right)^{\frac{1}{4}}(384(24)^{n-2}\alpha
n(n-1)(g-1))^{\frac{3}{4}}(\gamma
g)^{\frac{1}{4}(12n-1)}\\\label{39}&\times&\left(\frac{12n-12ng+9g-6}{4\gamma
g(n_{s}-1)}\right)^{\frac{(12ng-12n-g-2)}{4g}}.
\end{eqnarray}
\begin{figure}
\epsfig{file=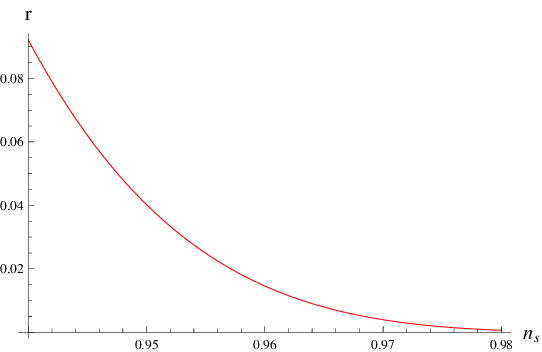, width=0.5\linewidth}\epsfig{file=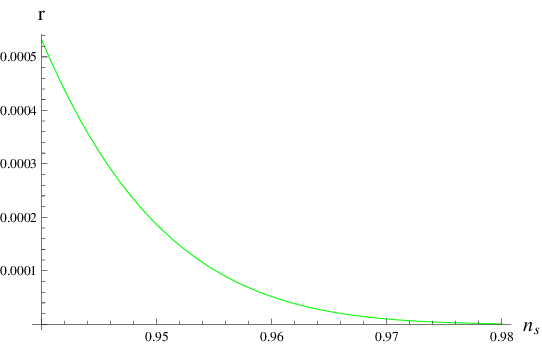,
width=0.5\linewidth}\caption{$r$ versus $n_{s}$ for
$\gamma=1,~g=0.5,~\alpha=-1\times10^{-3}, C=70$ and $\chi_{0}\propto
C^{\frac{1}{6}}$. The left plot is for $n=1.1$ and right for
$n=1.5$.}
\end{figure}
\begin{figure}
\epsfig{file=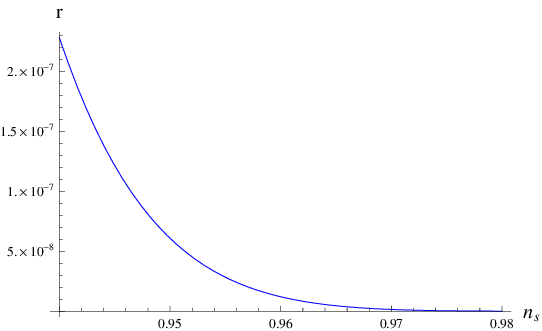, width=0.5\linewidth}\epsfig{file=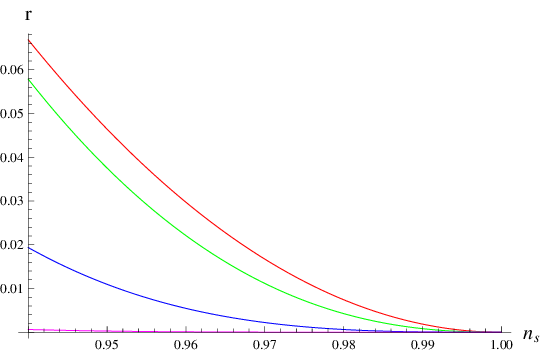,
width=0.5\linewidth}\caption{$r$ versus $n_{s}$ for
$\gamma=1,~\alpha=-1\times10^{-3}, C=70$ and $\chi_{0}\propto
C^{\frac{1}{6}}$. The left plot is for $n=2$ with $g=0.5$ and right
for $g=0.8,~n=1.5$ (red), $n=2$ (green), $n=3$ (blue) and $n=5$
(magenta).}
\end{figure}
The graphical behavior of $r$ versus $n_{s}$ is given in Figures
\textbf{2} and \textbf{3}. The model is compatible with Planck data
for the values $n=1.1,~1.5$ as shown in Figure \textbf{2}. The left
panel of Figure \textbf{3} gives incompatible value of $r$ as it
gets very small for $n=2$ and hence can be neglected. In the right
plot, $n_{s}$ lies in the region where $r<0.11$ for $n=1.5,~2,~3$
whereas the observational value of $r$ is inconsistent for $n=5$.

\subsection{Case II: $\chi=\chi_{*}\frac{T^3}{\varphi^2}$}

In this case, the scalar field and Hubble parameter are
\begin{equation}\label{40}
\varphi=\varphi_{0}\exp(\Delta_{2}
t^{\frac{3}{8}g-\frac{n}{2}(1-g)+\frac{1}{2}}), \quad H=\gamma g
\left(\frac{\ln\varphi-\ln\varphi_{0}}{\Delta_{2}}\right)^
{\frac{(g-1)}{\frac{3}{8}g-\frac{n}{2}(1-g)+\frac{1}{2}}},
\end{equation}
where $\Delta_{2}=\frac{8(4C\gamma
g)^{\frac{3}{8}}}{\chi_{*}^{\frac{1}{2}}(3g-4n(1-g)+4)}
[6(384)(24)^{n-2}\alpha n(n-1)(g-1)(\gamma g)^{4n}]^{\frac{1}{8}}$
with $\alpha<0$. The slow-roll parameters lead to
\begin{eqnarray}\label{41}
\varepsilon&=&\left(\frac{1-g}{\gamma
g}\right)\left(\frac{\ln\varphi-\ln\varphi_{0}}{\Delta_{2}}\right)^
{\frac{-g}{\frac{3}{8}g-\frac{n}{2}(1-g)+\frac{1}{2}}},\\\label{42}
\eta&=&\left(\frac{2-g}{\gamma
g}\right)\left(\frac{\ln\varphi-\ln\varphi_{0}}{\Delta_{2}}\right)^
{\frac{-g}{\frac{3}{8}g-\frac{n}{2}(1-g)+\frac{1}{2}}}.
\end{eqnarray}
The corresponding radiation density can be described as
\begin{equation}\nonumber
\rho_{r}=\frac{3}{2}(384)(24)^{n-2}\alpha n(n-1)(\gamma
g)^{4n-1}(g-1)\left(\frac{\ln\varphi-\ln\varphi_{0}}{\Delta_{2}}\right)
^{\frac{4ng-4n-g}{\frac{3}{8}g-\frac{n}{2}(1-g)+\frac{1}{2}}}.
\end{equation}
The number of e-folds between $\varphi_{0}$ and $\varphi_{i}$ is
given as
\begin{equation}\label{43}
\mathcal{N}=\gamma\left[\left(\frac{\ln\varphi-\ln\varphi_{0}}{\Delta_{2}}\right)^
{\frac{g}{\frac{3}{8}g-\frac{n}{2}(1-g)+\frac{1}{2}}}-\left(\frac{\ln\varphi_{i}-
\ln\varphi_{0}}{\Delta_{2}}\right)^{\frac{g}{\frac{3}{8}g-\frac{n}{2}(1-g)
+\frac{1}{2}}}\right],
\end{equation}
where
\begin{equation}\nonumber
\varphi_{i}=\varphi_{0}\exp\left[\Delta_{2}\left(\frac{1-g}{\gamma
g}\right)^{\frac{\frac{3}{8}g-\frac{n}{2}(1-g)+\frac{1}{2}}{g}}\right].
\end{equation}
Inserting the value of $\varphi_{i}$ in Eq.(\ref{43}), we have
\begin{equation}\label{44}
\varphi=\varphi_{0}\exp\left[\Delta_{2}\left(\frac{\mathcal{N}}{\gamma}
+\frac{1-g}{\gamma g}\right)^{\frac{\frac{3}{8}g
-\frac{n}{2}(1-g)+\frac{1}{2}}{g}}\right].
\end{equation}
The scalar power spectrum and spectral index have the following
expressions in terms of $\mathcal{N}$ as
\begin{eqnarray}\nonumber
\mathcal{P}_{s}&=&\left(\frac{\chi_{*}^{3}}{36(4\pi)^3}\right)^{\frac{1}{2}}
\left[\left(\frac{3}{2C}\right)^{11}\left(384(24)^{n-2}\alpha
n(n-1)(\gamma
g)^{4n+3}(g-1)\right)^3\right]^{\frac{1}{8}}\\\nonumber&\times&\varphi_{0}^{-3}
\exp\left[-3\Delta_{2}\left(\frac{\mathcal{N}}{\gamma}+\frac{1-g}{\gamma
g}\right)^{\frac{1}{g}(\frac{3}{8}g-\frac{n}{2}(1-g)+\frac{1}{2})}\right]
\\\nonumber&\times&\left(\frac{\mathcal{N}}{\gamma}+\frac{1-g}{\gamma
g}\right)^{\frac{76ng-76n-23g+20}{8g}},\\\label{45} n_{s}&=&
1-\frac{76ng-76n-23g+20}{8\gamma g}\left(\frac{\mathcal{N}}
{\gamma}+\frac{1-g}{\gamma g}\right)^{-1}.
\end{eqnarray}
Figure \textbf{4} represents negative values of $\mathcal{N}$ and
hence the required amount of inflation cannot be obtained for
variable dissipation factor. Consequently, the observational
parameter is not consistent with Planck data.
\begin{figure}
\epsfig{file=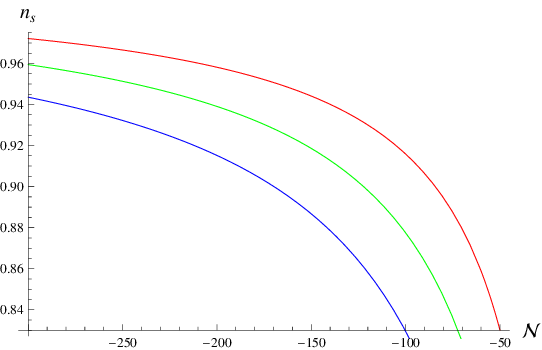, width=0.5\linewidth}\epsfig{file=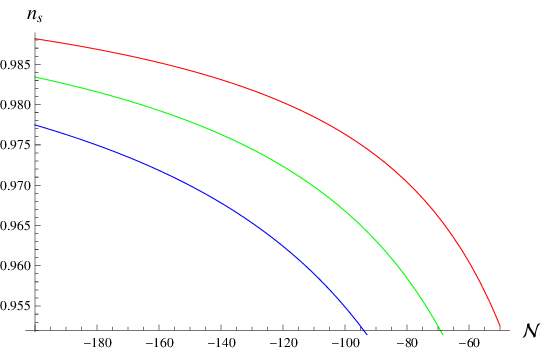,
width=0.5\linewidth}\caption{$n_{s}$ versus $\mathcal{N}$ for
$\gamma=1$. The left plot is for $g=0.5,~n=1.1$ (red), $n=1.5$
(green) and $n=2$ (blue) and right for $g=0.8$.}
\end{figure}

\section{Logamediate Inflation}

In this section, we study the dynamics of warm logamediate inflation
for the power-law model (\ref{27}). The logamediate inflation is
motivated by applying weak general conditions on the indefinitely
expanding cosmological models. For this regime, the scale factor has
the form \cite{13}
\begin{equation}\label{47}
a(t)=a_{0}\exp(\lambda[\ln t]^h),\quad\lambda>0, \quad h>1.
\end{equation}
For $h = 1$, this model is converted to power-law inflation
$(a(t)=a_{0}t^{\lambda},~\lambda > 1)$. The corresponding number of
e-folds reads
\begin{equation}\label{48}
\mathcal{N}=\lambda[(\ln t)^h-(\ln t_{i})^h].
\end{equation}
In the following, we evaluate inflationary parameters for the
constant as well as variable dissipation factor.

\subsection{Case I: $\chi=\chi_{0}$}

The solution for the inflaton is obtained using Eqs.(\ref{27}) and
(\ref{47}) in (\ref{18}) as follows
\begin{equation}\label{49}
\varphi=\varphi_{0}+\Delta_{3}\Omega(t),
\end{equation}
where
\begin{eqnarray}\nonumber
\Delta_{3}=-\left[\frac{6}{\chi_{0}}(-384(24)^{n-2}\alpha
n(n-1))\right]^{\frac{1}{2}}(\lambda
h)^{2n}\left[-\left(\frac{1}{2}-2n\right)\right]^{-2(h-1)n-1},
\end{eqnarray}
and $\Omega(t)=\Gamma\left[2(h-1)n+1,-(\frac{1}{2}-2n)\ln t\right]$
is an incomplete gamma function. The Hubble parameter is
\begin{equation}\label{50}
H=\lambda h
\frac{\left[\ln\left(\Omega^{-1}\left(\frac{\varphi-\varphi_{0}}
{\Delta_{3}}\right)\right)\right]^{h-1}}{\Omega^{-1}
\left(\frac{\varphi-\varphi_{0}}{\Delta_{3}}\right)}.
\end{equation}
Using the above equation in Eq.(\ref{16}), the slow-roll parameters
are
\begin{equation}\label{51}
\varepsilon=\frac{1}{\lambda h}\left[\ln\Omega^{-1}
\left(\frac{\varphi-\varphi_{0}}{\Delta_{3}}\right)\right]^{1-h},
\quad \eta=\frac{2}{\lambda h}\left[\ln\Omega^{-1}
\left(\frac{\varphi-\varphi_{0}}{\Delta_{3}}\right)\right]^{1-h}.
\end{equation}
The number of e-folds is
\begin{equation}\nonumber
\mathcal{N}=\lambda\left[\left(\ln\Omega^{-1}\left(\frac{\varphi
-\varphi_{0}}{\Delta_{3}}\right)\right)^{h}-\left(\ln\Omega^{-1}
\left(\frac{\varphi_{i}-\varphi_{0}}{\Delta_{3}}\right)\right)^{h}\right].
\end{equation}
Evaluating the value of $\varphi_{i}$ at $\varepsilon=1$ and
inserting in the above equation, it follows that
\begin{equation}\label{52}
\mathcal{N}=\lambda\left[\left(\ln\Omega^{-1}\left(\frac{\varphi
-\varphi_{0}}{\Delta_{3}}\right)\right)^{h}-(\lambda
h)^{\frac{h}{1-h}}\right],
\end{equation}
which can also be written as
\begin{equation}\label{53}
\varphi=\varphi_{0}+\Delta_{3}\Omega\left[\exp\left\{\left
(\frac{\mathcal{N}}{\lambda}+(\lambda h)^{\frac{h}{1-h}}\right)
^{\frac{1}{h}}\right\}\right].
\end{equation}
\begin{figure}
\epsfig{file=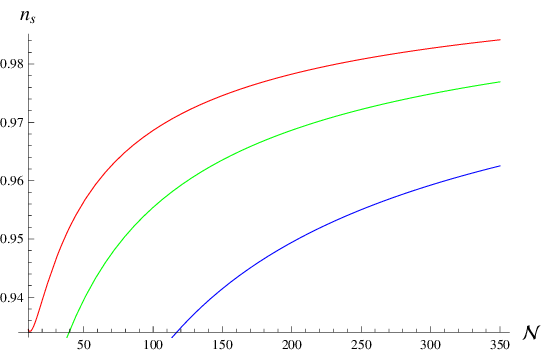, width=0.5\linewidth}\epsfig{file=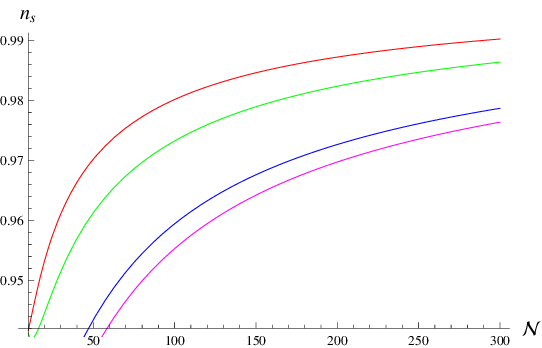,
width=0.5\linewidth}\caption{$n_{s}$ versus $\mathcal{N}$. (Left)
$\lambda=1,~h=3,n=1.5$ (red), $n=2$ (green) and $n=3$ (blue).
(Right) $\lambda=0.01,~h=5,n=1.5$ (red), $n=2$ (green), $n=3$ (blue)
and $n=3.3$ (magenta).}
\end{figure}

The scalar and tensor power spectra in terms of $\mathcal{N}$ are
\begin{eqnarray}\nonumber
\mathcal{P}_{s}&=&\left(\frac{\chi_{0}^3}{36(4\pi)^3}\right)
^{\frac{1}{2}}\left(\frac{3}{2C}\right)^{\frac{1}{4}}[-384(24)^{n-2}\alpha
n(n-1)]^{\frac{-3}{4}}(\lambda
h)^{\frac{3}{4}(3-4n)}\\\nonumber&\times&
\left[\exp\left\{\left(\frac{\mathcal{N}}{\lambda}+(\lambda
h)^{\frac{h}{1-h}}\right)^{\frac{1}{h}}\right\}\right]
^{3n-\frac{3}{2}}\left(\frac{\mathcal{N}}{\lambda}+(\lambda
h)^{\frac{h}{1-h}}\right)^{\frac{3}{4h}(h-1)(3-4n)},\\\nonumber
\mathcal{P}_{T}&=&\frac{2\kappa^2}{\pi^2}(\lambda
h)^2\left(\frac{\mathcal{N}}{\lambda}+(\lambda
h)^{\frac{h}{1-h}}\right)^{\frac{2}{h}(h-1)}
\exp\left[-2\left(\frac{\mathcal{N}}{\lambda}+(\lambda
h)^{\frac{h}{1-h}}\right)^{\frac{1}{h}}\right].
\end{eqnarray}
Equations (\ref{22}) and (\ref{25}) lead to the scalar and tensor
spectral indices, respectively as
\begin{eqnarray}\nonumber
n_{s}&=&1-\frac{1}{\lambda
h}\left(3n-\frac{3}{2}\right)\left(\frac{\mathcal{N}}{\lambda}+(\lambda
h)^{\frac{h}{1-h}}\right)^{\frac{1}{h}-1}-\frac{3}{4\lambda
h}(h-1)(3-4n)\\\label{54}&\times&\left(\frac{\mathcal{N}}{\lambda}+(\lambda
h)^{\frac{h}{1-h}}\right)^{-1},\\\nonumber n_{T}&=&-\frac{2}{\lambda
h}\left(\frac{\mathcal{N}}{\lambda}+(\lambda
h)^{\frac{h}{1-h}}\right)^{\frac{1-h}{h}}.
\end{eqnarray}
The increasing behavior of $n_{s}$ versus $\mathcal{N}$ is shown in
Figure \textbf{5}. In the left plot, for $n=1.5,~2,~3$, the
corresponding values of $\mathcal{N}$ are $63,~125,~313$. The number
of e-folds $29,~47,~105$ and $126$ are obtained for $n=1.5,~2,~3$
and $3.3$ in the right plot. In terms of $\mathcal{N}$, the
corresponding tensor-scalar ratio is
\begin{eqnarray}\nonumber
r&=&\left(\frac{144\kappa^4(4\pi)^3}{\chi_{0}^3\pi^{4}}\right)
^{\frac{1}{2}}[-384(24)^{n-2}\alpha n(n-1)]^{\frac{3}{4}}
\left(\frac{2C}{3}\right)^{\frac{1}{4}}(\lambda
h)^{3n-\frac{1}{4}}\\\nonumber&\times&\exp\left[\left(-3n-\frac{1}{2}
\right)\left(\frac{\mathcal{N}}{\lambda}+(\lambda
h)^{\frac{h}{1-h}}\right)^{\frac{1}{h}}\right]\left(\frac{\mathcal{N}}
{\lambda}+(\lambda h)^{\frac{h}{1-h}}\right)
^{\frac{1}{h}(h-1)\left(3n-\frac{1}{4}\right)}.\\\label{55}
\end{eqnarray}
Figure \textbf{6} shows the parametric plot of Eqs.(\ref{54}) and
(\ref{55}). The left panel shows that compatible results are
obtained for $n=1.5$ and $n=2$ while the observational parameters
are in agreement for $n=2$ and $3$ in the right plot.
\begin{figure}
\epsfig{file=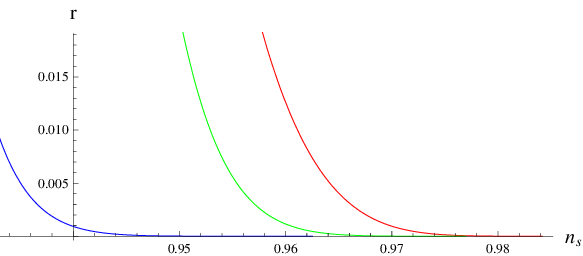,
width=0.5\linewidth}\quad\quad\epsfig{file=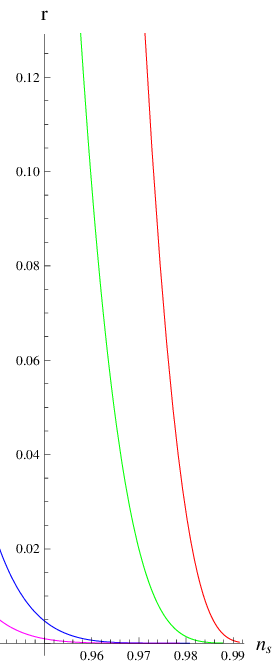,
width=0.2\linewidth}\caption{$r$ versus $n_{s}$ for
$\alpha=-1\times10^{-3},~C=70,~\chi_{0}\propto C^{\frac{1}{6}},
~n=1.5$ (red), $n=2$ (green) and $n=3$ (blue). The left plot is for
$\lambda=1$ with $h=3$ and right for $\lambda=0.01,~h=5$ and $n=3.3$
(magenta).}
\end{figure}

\subsection{Case II: $\chi=\chi_{*}\frac{T^3}{\varphi^2}$}

In this case, the scalar field is found to be
\begin{equation}\label{56}
\varphi=\varphi_{0}\exp(\Delta_{4}\Xi(t)),
\end{equation}
where
\begin{eqnarray}\nonumber
\Delta_{4}&=&-\left(\frac{6}{\chi_{*}}\right)^{\frac{1}{2}}
\left(\frac{2C}{3}\right)^{\frac{3}{8}}[-384(24)^{n-2}\alpha
n(n-1)]^{\frac{1}{8}}(\lambda
h)^{\frac{1}{8}(4n+3)}\\\nonumber&\times&
\left(\frac{n-1}{2}\right)^{\frac{-1}{8}(h-1)(4n+3)-1},\quad
\alpha<0,
\end{eqnarray}
and $\Xi(t)=\Gamma\left[\frac{1}{8}(h-1)
(4n+3)+1,\frac{1}{2}(n-1)\ln t\right]$ is an incomplete gamma
function. The Hubble parameter becomes
\begin{equation}\label{57}
H=\frac{\lambda h\left[\ln\Xi^{-1}
\left(\frac{\ln\varphi-\ln\varphi_{0}}{\Delta_{4}}\right)\right]^{h-1}}
{\Xi^{-1}\left(\frac{\ln\varphi-\ln\varphi_{0}}{\Delta_{4}}\right)},
\end{equation}
and the corresponding $(\varepsilon,\eta)$ parameters are
\begin{equation}\label{58}
\varepsilon=\frac{1}{\lambda
h}\left[\ln\Xi^{-1}\left(\frac{\ln\varphi-\ln\varphi_{0}}{\Delta_{4}}
\right)\right]^{1-h},\quad\eta=\frac{2}{\lambda
h}\left[\ln\Xi^{-1}\left(\frac{\ln\varphi-\ln\varphi_{0}}{\Delta_{4}}
\right)\right]^{1-h}.
\end{equation}
The number of e-folds is given as
\begin{eqnarray}\nonumber
\mathcal{N}&=&\lambda\left[\left(\ln\Xi^{-1}\left(\frac{\ln\varphi-\ln\varphi_{0}}
{\Delta_{4}}\right)\right)^{h}-\left(\ln\Xi^{-1}\left(\frac{\ln\varphi_{1}
-\ln\varphi_{0}}{\Delta_{4}}\right)\right)^{h}\right],\\\label{59}&=&
\lambda\left[\left(\ln\Xi^{-1}\left(\frac{\ln\varphi-\ln\varphi_{0}}
{\Delta_{4}}\right)\right)^{h}-(\lambda h)^{\frac{h}{1-h}}\right],
\end{eqnarray}
while the scalar field in terms of $\mathcal{N}$ is expressed as
\begin{equation}\label{60}
\varphi=\varphi_{0}\exp\left[\Delta_{4}\Xi\left[\exp\left
\{\left(\frac{\mathcal{N}}{\lambda}+(\lambda
h)^{\frac{h}{1-h}}\right)^{\frac{1}{h}}\right\}\right]\right].
\end{equation}
The spectrum parameters $\mathcal{P}_{s}$ and $\mathcal{P}_{T}$ are
\begin{eqnarray}\nonumber
\mathcal{P}_{s}&=&\left(\frac{\chi_{*}^3}{36(4\pi)^3}\right)^{\frac{1}{2}}
\left(\frac{3}{2C}\right)^{\frac{11}{8}}[-384(24)^{n-2}\alpha
n(n-1)]^{\frac{3}{8}}(\lambda h)^{\frac{3}{8}(4n+3)}
\varphi_{0}^{-3}\\\nonumber&\times&\exp\left[-3\Delta_{4}
\Xi\left(\exp\left\{\left(\frac{\mathcal{N}}{\lambda}+(\lambda
h)^{\frac{h}{1-h}}\right)^{\frac{1}{h}}\right\}\right)\right]
\exp\left[\frac{-3}{2}(n+1)\right.\\\nonumber&\times&
\left.\left(\frac{\mathcal{N}}{\lambda}+(\lambda
h)^{\frac{h}{1-h}}\right)^{\frac{1}{h}}\right]
\left(\frac{\mathcal{N}}{\lambda}+(\lambda
h)^{\frac{h}{1-h}}\right)^{\frac{3}{8h}(h-1)(4n+3)},\\\nonumber
\mathcal{P}_{T}&=&\frac{2\kappa^2}{\pi^2}(\lambda
h)^2\left(\frac{\mathcal{N}}{\lambda}+(\lambda
h)^{\frac{h}{1-h}}\right)^{\frac{2}{h}(h-1)}\exp\left[
-2\left(\frac{\mathcal{N}}{\lambda}+(\lambda
h)^{\frac{h}{1-h}}\right)^{\frac{1}{h}}\right].
\end{eqnarray}
\begin{figure}
\center\epsfig{file=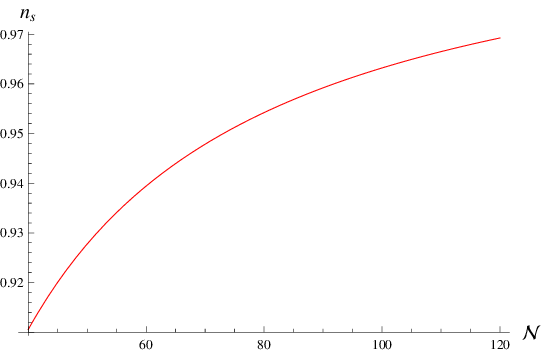, width=0.5\linewidth}\caption{$n_{s}$
versus $\mathcal{N}$ for $\lambda=0.01,~h=3,n=3$.}
\end{figure}

The corresponding spectral indices in terms of $\mathcal{N}$ are as
follows
\begin{eqnarray}\label{61}
n_{s}&=&1-\frac{3}{8h\lambda}(h-1)(4n+3)\left(\frac{\mathcal{N}}{\lambda}+(\lambda
h)^{\frac{h}{1-h}}\right)^{-1},\\\nonumber n_{T}&=&-\frac{2}{\lambda
h}\left(\frac{\mathcal{N}}{\lambda}+(\lambda
h)^{\frac{h}{1-h}}\right)^{\frac{1-h}{h}}.
\end{eqnarray}
Figure \textbf{7} yields the increasing behavior of $n_{s}$ with
respect to $\mathcal{N}$. The observational value $(n_{s}=0.96)$
corresponds to $\mathcal{N}=94$ which predicts the physical
compatibility of the isotropic model. Using Eqs.(\ref{19}),
(\ref{27}), (\ref{47}) and (\ref{60}) in Eq.(\ref{26}), we get the
expression for $r$ as
\begin{eqnarray}\nonumber
r&=&\left(\frac{144\kappa^4(4\pi)^3}{\chi_{*}^3\pi^{4}}\right)
^{\frac{1}{2}}\left(\frac{2C}{3}\right)^{\frac{11}{8}}[-384(24^{n-2}\alpha
n(n-1))]^{\frac{-3}{8}}(\lambda
h)^{\frac{1}{8}(-12n+7)}\\\nonumber&\times&\varphi_{0}^{3}
\exp\left[3\Delta_{4}\Xi\left(\exp\left\{\left(\frac{\mathcal{N}}{\lambda}
+(\lambda
h)^{\frac{h}{1-h}}\right)^{\frac{1}{h}}\right\}\right)\right]
\exp\left[\frac{1}{8}(12n-4)\right.\\\nonumber&
\times&\left.\left(\frac{\mathcal{N}}{\lambda}+(\lambda
h)^{\frac{h}{1-h}}\right)^{\frac{1}{h}}\right]\left(\frac{\mathcal{N}}{\lambda}
+(\lambda h)^{\frac{h}{1-h}}\right)^{\frac{1}{8h}(h-1)(-12n+7)},
\end{eqnarray}
which can also be written in the form of $n_{s}$ as
\begin{eqnarray}\nonumber
r&=&\left(\frac{144\kappa^4(4\pi)^3}{\chi_{*}^3\pi^{4}}\right)
^{\frac{1}{2}}\left(\frac{2C}{3}\right)^{\frac{11}{8}}[-384(24^{n-2}\alpha
n(n-1))]^{\frac{-3}{8}}(\lambda
h)^{\frac{1}{8}(-12n+7)}\\\nonumber&\times&\varphi_{0}^{3}
\exp\left[3\Delta_{4}\Xi\left(\exp\left\{\left(\frac{3(h-1)(4n+3)}{8\lambda
h(1-n_{s})}\right)^{\frac{1}{h}}\right\}\right)\right]
\exp\left[\frac{1}{8}(12n-4)\right.\\\label{62}&\times&\left.\left(\frac{3(h-1)(4n+3)}{8\lambda
h(1-n_{s})}\right)^{\frac{1}{h}}\right]\left(\frac{3(h-1)(4n+3)}{8\lambda
h(1-n_{s})}\right)^{\frac{1}{8h}(h-1)(-12n+7)}.
\end{eqnarray}
The graphical behavior of $r$ versus $n_{s}$ is shown in Figure
\textbf{8}. Three different choices of dissipation factor gives
compatible results for $n=3$ with Planck data.
\begin{figure}
\center\epsfig{file=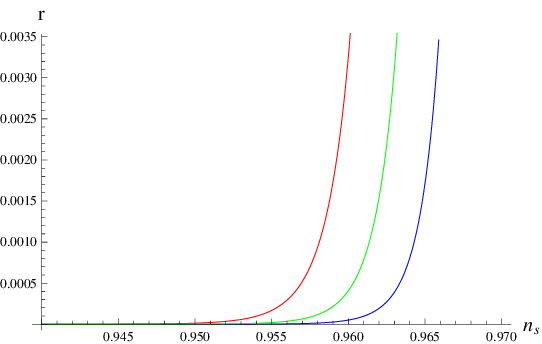, width=0.5\linewidth}\caption{$r$
versus $n_{s}$ for $n=3,~\lambda=0.01,~h=3,~C=70,~\varphi_{0}\propto
C^{\frac{-1}{12}},~\alpha=-1\times10^{-3},~\chi_{*}=0.25$ (red),
$1$(green) and $4$(blue).}
\end{figure}

\section{Conclusions}

In this paper, we have explored warm inflation for FRW universe
model in the background of $f(\mathcal{G})$ gravity. In warm
inflation, the interactions between scalar and other fields are
taken into account which give the dissipation term. We have
formulated the conservation equation, slow-roll parameters
$(\varepsilon,\eta)$, scalar and tensor power spectra
$(\mathcal{P}_{s},\mathcal{P}_{T})$, spectral indices
$(n_{s},n_{T})$ and tensor-scalar ratio using the field equations
under the slow-roll approximations.

We have investigated intermediate and logamediate inflationary eras
for strong dissipative regime. In each case, we have assumed two
specific forms of dissipation factor (a positive constant and a
function of scalar field) and a power-law model of $f(\mathcal{G})$
gravity and have evaluated all the above mentioned parameters. The
trajectories of $\mathcal{N}$ and $r$ versus $n_{s}$ have been
plotted to check the compatibility of the model with observational
Planck data in each case. The results are summarized as follows.
\begin{itemize}
\item In intermediate regime, for $g=0.5$ and $\gamma=1$, the number of
e-folds increases by increasing the values of $n$ while the
corresponding $r-n_{s}$ trajectories lead to physical compatible
range $1<n<2$. When $g=0.8$, it is found that $n$ lies between 1 and
5 as shown in Figure \textbf{3} for constant dissipation factor.
\item For the variable dissipation factor in intermediate epoch, the model is
inconsistent with observational data.
\item In logamediate inflationary era, $n\leq2$ gives consistent
results for $h=3$ and $\lambda=1$. For $h=5$ and $\lambda=0.01$, the
physical acceptable range is $2\leq n<3.3$ as shown in Figures
\textbf{5} and \textbf{6}.
\item In the second case of logamediate era, the effect of
dissipation coefficient $\chi_{*}$ is examined for
$n=3,~\lambda=0.01$ and $h=3$. It is shown in Figure \textbf{8} that
the results are compatible with observational data for all chosen
values of $\chi_{*}$ .
\end{itemize}
Finally, we conclude that power-law model is consistent with Planck
data except for the variable dissipation factor in intermediate
inflationary regime.

\end{document}